# Deep-time consistency in proteome elemental composition across cellular and viral life

L. Felipe Benites[1], Louie Slocombe[1], Sara Imari Walker[1,2,3*]

[1]Beyond Center for Fundamental Concepts in Science, Arizona State University, Tempe AZ USA

[2]School of Earth and Space Exploration, Arizona State University, Tempe AZ USA

[3]Santa Fe Institute, Santa Fe, NM USA

*Author for correspondence: sara.i.walker@asu.edu.

**Abstract**

Proteins are constructed from a limited alphabet of ~20 amino acids, yet the origins and selection of this specific alphabet remain unresolved. One largely overlooked aspect in protein evolution is whether elemental composition constrains the range of viable proteomes. Here, we analyzed the elemental composition of thousands of proteomes spanning cellular domains and viral realms. Despite extensive evolutionary divergence and orders-of-magnitude variation in proteome size and gene content, proteomes exhibited strikingly consistent elemental composition. This consistency was substantially more constrained than amino acid frequencies or physicochemical properties and was not explained by evolutionary relatedness, biological function, or amino acid usage alone. Viral proteomes across the major realms occupied largely the same elemental composition space observed in cellular organisms despite the absence of a single viral common ancestor, suggesting that common biochemical constraints shape proteome organization across life. To investigate the evolutionary origins of this pattern, we compared modern proteomes with multiple independent reconstructions of the Last Universal Common Ancestor (LUCA) and with synthetic reduced-alphabet proteomes generated using proposed primordial amino acid alphabets. LUCA-derived proteomes occupied the same constrained elemental composition space observed in modern Bacteria and Archaea, whereas reduced primordial-like alphabets systematically generated alternative elemental regimes outside the modern range despite retaining substantial detectable sequence similarity to extant proteins. Reduced alphabets also disrupted fold space and reorganized relationships between elemental composition and predicted protein structural organization. Together, our results suggest that constrained proteome elemental composition represents a fundamental organizational property of biological proteomes that emerged early in evolution and may have contributed to the selection and stabilization of the modern amino acid alphabet.

**Main**

Proteins are large macromolecules composed of linear chains of amino acids. They are essential for all known forms of life on Earth, including viruses. Twenty of the twenty-two coded amino acids are universally utilized, but two (Selenocysteine and Pyrrolysine) are rarer (Böck et al., 1991; Hao et al., 2002). However, more than 100 amino acids have been identified in nature, notably in abiotic examples like meteorites (Ibba et al. 2001) indicating these could have been present at the origin of life. Furthermore, thousands of amino acids are theoretically possible based on structural exploration of chemical space (Ilardo et al. 2015). These can be classified according to core functional groups as α-, β-, γ-, δ-, with most having one or more enantiomeric mirror image forms (L or D). However, the genetically encoded amino acids are exclusively alpha (α)-levorotatory (L) type, reflecting a universal homochiral stereochemical bias. The origin and early evolution selecting this set, with its specific properties, among the potential vast landscape of other chemical possibilities remains uncertain (Blackmond, 2019).

Many hypotheses have been explored to explain the selection of the amino acid alphabet (Lu and Freeland 2006). These range from early chemical selection on physicochemical properties of amino acids, such as the elimination of poorly soluble aliphatic and basic amino acids with limited secondary structures, to potentially later-stage foldability constraints that could have narrowed the amino acid alphabet to what it is today (Makarov et al. 2023). The coded set differs in structure and in elemental content, with less nitrogen and more sulfur and oxygen, as compared to potentially viable alternative sets (including stable isomers and near-isomers of the coded alphabet (Brown et al. 2023). The elemental differences could be indicative of elemental compositional constraints that could have also shaped the emergence and evolution of proteomes in parallel to other physicochemical properties. However, the role of elemental composition has been largely overlooked in studies seeking to explain the emergence and persistence of the coded set (Baudouin-Cornu et al. 2004).

Elements play a critical role in the organization of life (Remick and Helmann; 2023). Proteins consist of just six elements: carbon (C), hydrogen (H), nitrogen (N), oxygen (O), sulfur (S), and selenium (Se) (Asplund et al. 2006; Brown et al. 2023). Selection for resource conservation was proposed to influence the structure of the standard genetic code, specifically for N and C content (Shenhav and Zeevi, 2020), although these conclusions were later challenged (Rozhoňová and Payne, 2021). Prior proteomic studies reported variation in sulfur and carbon usage associated with metabolic function and nutrient limitation in proteins from Bacteria and Eukaryota (Baudouin-Cornu et al. 2001, 2004), raising the possibility that protein elemental composition may primarily reflect ecological or metabolic adaptation rather than universal constraints. Across different scales, consistent elemental relationships between environmental availability and organismal composition have been identified. A classic example is the Redfield ratio in oceanic phytoplankton and seawater, where carbon, nitrogen, and phosphorus occur at an approximately constant ratio of 106C:16N:1P (Redfield, 1934). Given the central importance of elemental stoichiometry to living systems, we investigated here if proteomes themselves are also subject to universal elemental constraints. We hypothesized that such constraints contributed to shaping the evolution of the amino acid alphabet and the global organization of modern proteomes.

To explore this hypothesis, we analyze the collective properties of proteomes (here defined as the full set of proteins predicted in a genome) as ensembles of amino acid chains. Here we analyzed the elemental, and compositional features of proteomes and their relationships across a large-scale dataset of all major groups of life, including viruses. Remarkably, despite variations in protein number, length, elemental quantity and amino acid frequencies, we observed a strikingly consistent elemental composition across global proteomes. This consistency persists despite deep evolutionary divergence, phylogenetic unrelatedness, and the vast range of encoded biological functions that evolved since the time of LUCA.

## Results and discussion

### The global elemental composition of cellular and viral proteomes is consistent despite large differences in proteomes and evolutionary history

To characterize proteome elemental composition, we quantified carbon (C), hydrogen (H), nitrogen (N), oxygen (O), sulfur (S), and selenium (Se) across high quality cellular (n = 6,964) and viral (n = 489,025) proteomes and normalized counts by total atoms per proteome. Total proteome sizes and atom abundances varied by several orders of magnitude, ranging from one to two proteins and 2.67×10³ atoms in Ribozyviria (viruses) to 117,502 proteins and 1.25×10⁹ atoms in Eukaryota (cells) **(Fig. 1 A-B) (Supplementary information 01).** Despite this variation, elemental composition remained highly constrained across cellular and viral groups, with minimal variations. For example, C remained centered around ~31–32% across most proteomes, whereas H remained ~50%, N ~8–9%, O ~9–10%, S ~0.1–0.3%, and Se was nearly absent across all groups, with the exception of cellular organisms and one single instance in viruses (Varidnaviria). To test whether these patterns reflect biological constraints, we compared real proteomes to simulated random proteomes generated with uniform amino acid frequencies and parameters based on average bacterial proteomes. Random proteomes occupied a distinct compositional profile with high C, S, and Se together with reduced H and O relative to real proteomes. While most cellular and viral groups occupied overlapping compositional ranges. Ribozyviria represented the main exception, exhibiting lower C and higher N relative to other groups.

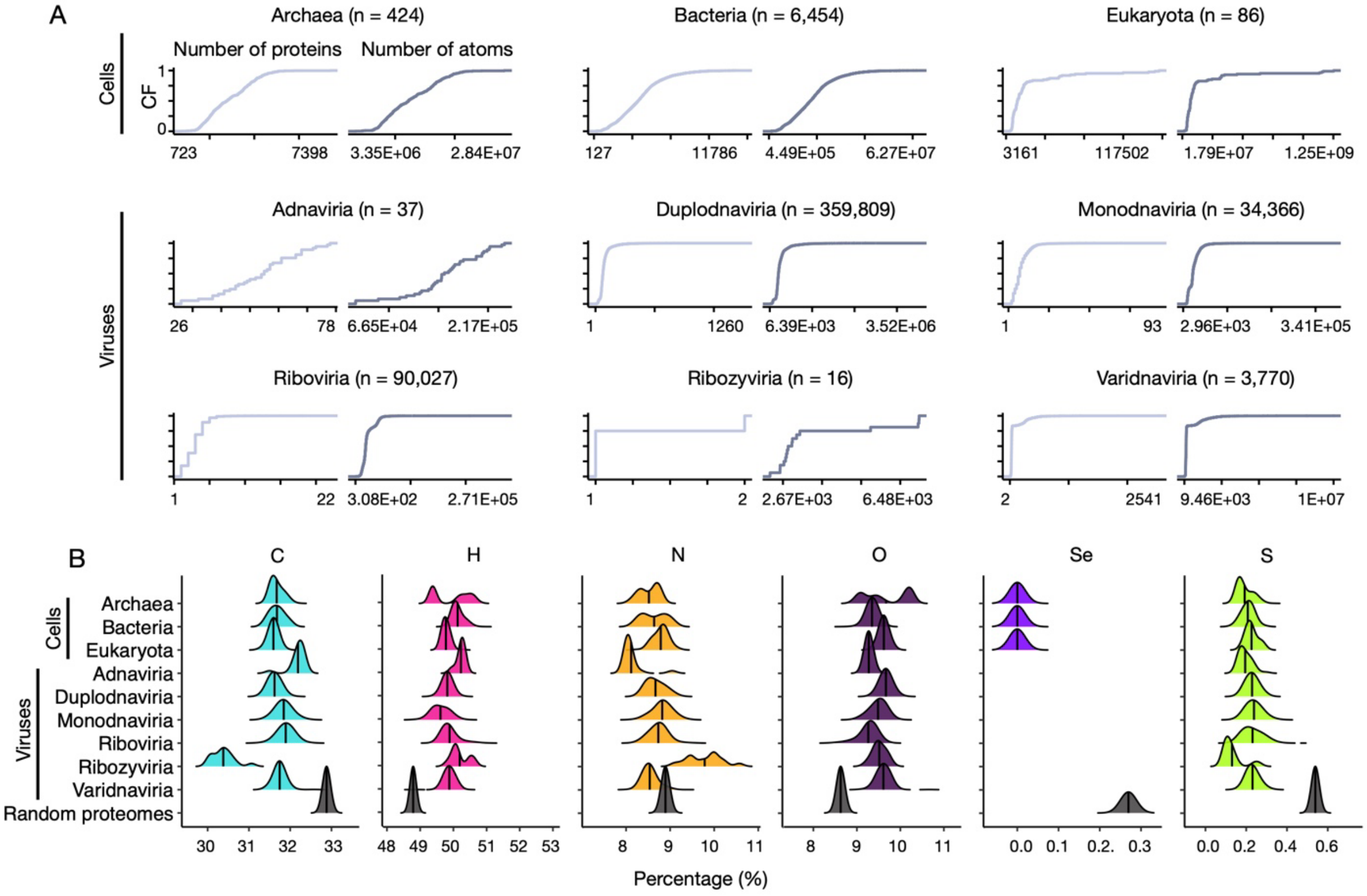

**Figure 1. Proteome size and elemental composition across cellular and viral life.** **(A)** Empirical cumulative distribution functions (ECDFs) showing the distribution of proteome size measured as number of proteins (light lavender blue) and total number of atoms (dark lavender blue) across cellular domains and viral realms. Proteomes are ordered from smallest to largest within each group. Numbers in parentheses indicate the number of proteomes analyzed. **(B)** Ridgeline density distributions of normalized proteome elemental composition for carbon (C),

hydrogen (H), nitrogen (N), oxygen (O), selenium (Se), and sulfur (S). Values represent the percentage contribution of each element relative to total proteome atoms. Colored distributions correspond to cellular domains and viral realms, whereas black distributions represent random proteomes generated as controls.

To compare elemental composition distributions among groups, we used pairwise Kolmogorov–Smirnov (KS) statistics. Most cellular and viral groups showed substantial distributional overlap. In contrast, random proteomes were consistently separated from biological proteomes ($D \approx 1$ across most elements), whereas Ribozyviria showed strong divergence primarily for C and N. PERMANOVA detected significant differences among groups ($R^2 = 0.72$, $p = 0.001$), although excluding random proteomes and Ribozyviria reduced the variance explained to $R^2 = 0.24$ **(Supplementary information 02)**. Principal component analysis (PCA) showed that elemental variation is low dimensional, with PC1 and PC2 explaining 28.3% and 25.6% of total variance, respectively **(Supplementary information 03, Supplementary Fig. 1)**. Most cellular and viral groups formed overlapping clusters, whereas Archaea exhibited a bimodal distribution, mainly in N, along PC1, associated with halophilic lineages Halobacteria and Nanohalarchaeota but also to *Methanopyrus kandleri* (Methanopyrales), which showed extreme intracellular salinity (Alexei I et al. 2002) (**Supplementary Fig. 2 A-B)**. In contrast, bacterial proteomes showed continuous variation primarily along a N–H gradient. Together, these results show that proteome elemental composition is highly consistent across cellular and viral life despite extreme differences in proteome size, and evolutionary history and with exception of some Archaea, ecological factors.

**Proteome elemental composition exhibits stronger global consistency than amino acid frequencies or physicochemical properties across cellular and viral life**

We next tested if proteome elemental composition is more constrained than amino acid frequencies and derived physicochemical properties, and if elemental composition can be directly explained by these features. To quantify relative divergence among proteomes, we calculated an inverted ratio (IR) metric based on normalized pairwise differences between feature vectors, where lower values indicate greater conservation and higher values indicate greater divergence. Discrete pairwise comparisons (IR) revealed that elemental composition shows markedly lower divergence across proteomes (median feature IR = 0.034, pairwise; 0.016, group median) than amino acid frequencies (0.247; 0.201) and amino acid properties (0.117; 0.078), indicating strong global constraint at the elemental level **(Supplementary information 04)**.

To evaluate whether this constraint is reflected in multivariate space, we quantified dispersion relative to group centroids. Elemental composition exhibited low dispersion (median distance = 0.127), compared to substantially higher dispersion for amino acid frequencies (0.851) and amino acid properties (1.512) **(Supplementary information 05)**. Although dispersion varied across groups, with higher variability in Archaea and Ribozyviria and lower variability in Eukaryota and Adnaviria, the relative ordering was consistent, with elemental composition always showing the lowest dispersion. We next tested the presence of lineage-specific structuring. Pairwise KS tests showed group differentiation, with moderate separation for elemental composition and amino acid frequencies (median $D \approx 0.42$–$0.43$), and stronger separation for amino acid properties (median $D \approx 0.55$). Individual comparisons ranged from near-complete overlap (e.g., Archaea vs Bacteria) to near-complete separation (e.g., Ribozyviria vs other groups), with C and N contributing most strongly to group differentiation **(Supplementary information 06)**.

Finally, we tested whether elemental composition can be explained by amino acid usage or properties. Global Spearman correlations between elemental composition and amino acid frequencies **(Fig. 2)** were weak (median $|r| = 0.12$; mean $|r| = 0.23$), and similarly weak for amino acid properties (median $|r| = 0.165$; mean $|r| = 0.239$), indicating that elemental composition is not directly determined by these features. Despite the weak global signal, specific element–feature relationships were highly recurrent across lineages, including strong associations between N and arginine (Arg - R) (e.g., in Archaea and Adnaviria), S and methionine (Met – M), C and aromatic residues, and H and aliphatic residues in most groups. Therefore, although amino acid usage and physicochemical properties influence elemental composition, they do not fully account for the global elemental consistency observed across proteomes. Together, these results show that proteome elemental composition exhibits stronger global consistency than amino acid frequencies or physicochemical properties, while still retaining lineage-specific variation not fully explained by these features.

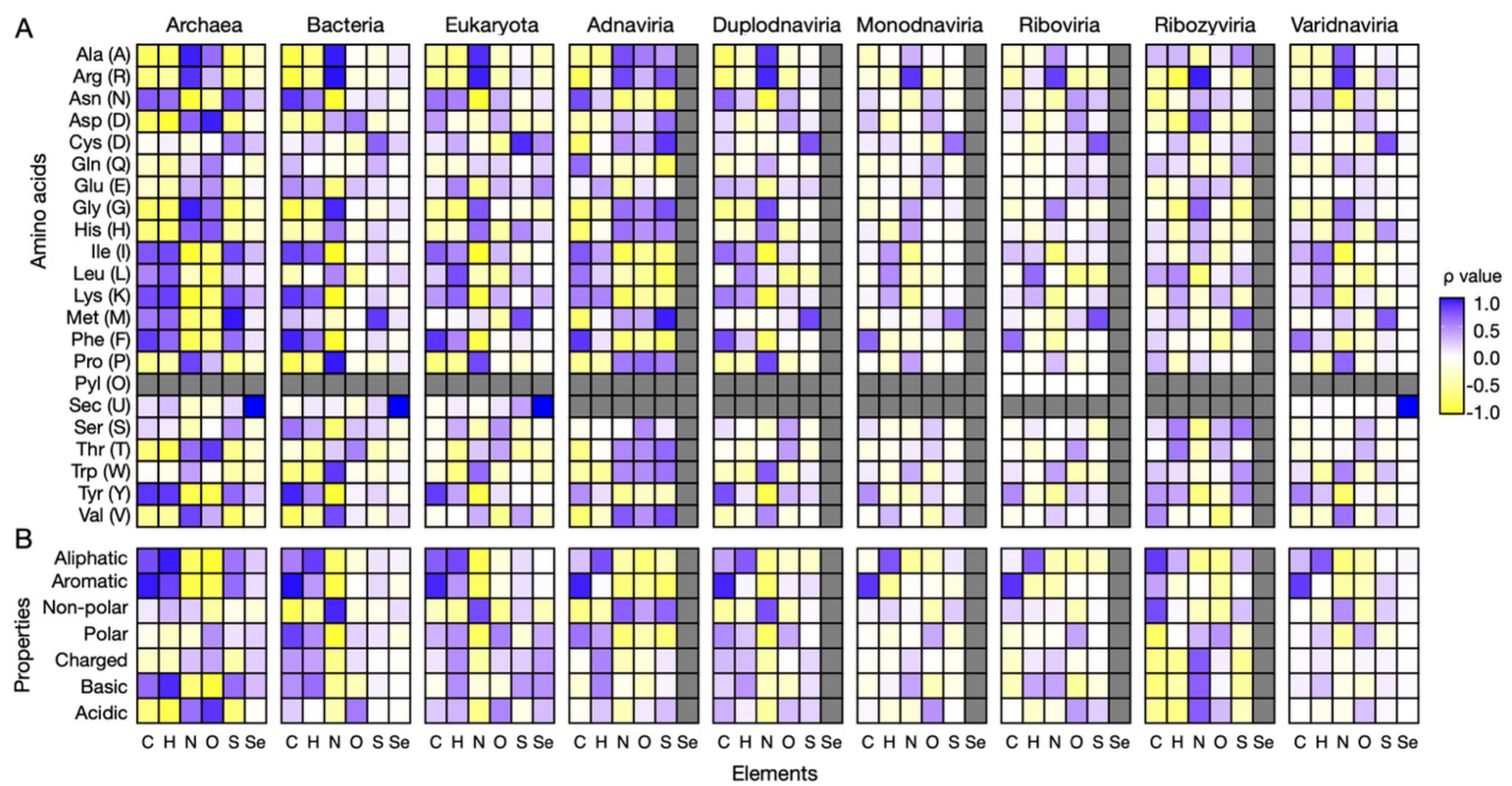

**Figure 2. Relationships between amino acid composition, physicochemical properties, and proteome elemental composition across cellular and viral life.** **(A)** Heatmap showing Spearman rank correlations (ρ) between amino acid frequencies and normalized elemental composition for carbon (C), hydrogen (H), nitrogen (N), oxygen (O), sulfur (S), and selenium (Se) across cellular domains and viral realms. Amino acids are shown on the y-axis using standard three and one-letter abbreviations. **(B)** Heatmap showing Spearman correlations between aggregated amino acid physicochemical property classes and elemental composition across the same groups. Properties include aliphatic, aromatic, non-polar, polar, charged, basic, and acidic residues. Colors indicate correlation strength and direction from negative (yellow) to positive (blue), with gray cells indicating absent values.

**The consistent elemental composition of cellular proteomes is conserved since the divergence from the Last Universal Common Ancestor (LUCA)**

We next asked whether the elemental consistency observed across proteomes reflects a modern feature or an ancestral property of proteomes. While viral realms lack a single common ancestor (Krupovic and Koonin, 2017; Harris and Hill, 2021), all cellular domains trace back to a hypothetical Last Universal Common Ancestor (LUCA). We therefore investigated whether this compositional signature might already have been established at the time of LUCA, which likely inhabited a markedly chemically different planetary environment characterized by higher $CO_2$ and sulfur availability (Weiss et al. 2016, 2018) and elevated hydrogen fluxes (Moody et al. 2024), parameters that could influence proteome composition (Knight et al. 2004). To address this, we compared independent LUCA-consensus proteomes, built from deep homologous sequences (PFAM -

Wehbi et al. 2024; COG – Crapitto et al. 2022; KO – Moody et al. 2024), including ancestrally reconstructed proteins (LUCA ASR - Wehbi et al. 2024) **(Supplementary information 07)**, with modern bacterial (n = 6,454) and archaeal (n = 424) proteomes. We found that all LUCA reconstructions fall within the range of inter-proteome Euclidean distances observed among modern proteomes (median = 0.504; range: 0.0006–3.23), that defines the elemental variance of extant cellular composition **(Table 1).**

| Metric | LUCA COG | LUCA KO | LUCA ASR | LUCA PFAM | Archaea (mean) | Bacteria (mean) |
|---|---|---|---|---|---|---|
| C (%) | 31.50 | 31.61 | 31.64 | 31.72 | 31.67 | 31.68 |
| H (%) | 50.12 | 50.25 | 50.42 | 50.16 | ~50.3 | ~50.2 |
| N (%) | 8.77 | 8.56 | 8.39 | 8.59 | 8.5 | 8.6 |
| O (%) | 9.40 | 9.34 | 9.32 | 9.31 | ~9.3 | ~9.3 |
| S (%) | 0.21 | 0.23 | 0.22 | 0.21 | 0.19 | 0.20 |
| Median distance (LUCA vs modern) | 0.394 | 0.414 | 0.518 | 0.382 | — | — |
| Median distance (modern vs modern) | — | — | — | — | 0.504 | 0.504 |

**Table 1. Elemental composition and compositional distances of LUCA-derived proteomes relative to modern cellular proteomes.** Elemental composition of LUCA consensus and ancestral sequence reconstructed (ASR) proteomes compared with mean archaeal and bacterial proteomes. Values for carbon (C), hydrogen (H), nitrogen (N), oxygen (O), and sulfur (S) are shown as percentages of total proteome atoms. LUCA-derived proteomes were reconstructed from conserved orthologous groups (COG, KO, PFAM) or ancestral sequence reconstruction (ASR). Median Euclidean distances between LUCA-derived and modern proteomes are shown together with the baseline median pairwise distance among modern bacterial and archaeal proteomes.

Consensus models derived from orthologous groups (PFAM, COG, KO) show slightly lower median distances to modern proteomes (0.38–0.41), consistent with representing conserved, averaged compositional profiles. In contrast, the LUCA ASR reconstruction (0.518) is at the center of the modern inter-proteome distance distribution, consistent with a plausible ancestral elemental composition. These results indicate that LUCA-derived proteomes occupy the shared elemental compositional space of modern Bacteria and Archaea without strong bias toward either domain, consistent with elemental constraints being established prior to their divergence.

**Pre-LUCA amino acid alphabets generate elemental compositions divergent from modern proteomes**

Because LUCA-derived proteomes already occupy the elemental composition space of modern cellular life, we next asked whether smaller primordial amino acid alphabets proposed for pre-LUCA evolution could also reproduce this compositional profile, or whether it emerged only after a gradual expansion of the encoded amino acid alphabet. Trifonov (2000) proposed a nine-residue primordial alphabet based on consensus amino acid emergence order, whereas Wehbi et al. (2024) reconstructed a different ten-residue early alphabet from inferred LUCA and pre-LUCA protein families. We therefore generated reduced alphabet "monster" proteomes (pre-LUCA and Trifonov monsters) through virtual mutagenesis of bacterial (wild type) proteomes by removing amino acids absent from each proposed alphabet while preserving the relative frequencies of retained residues **(Fig. 3 C)**. Although not intended as realistic evolutionary reconstructions, these synthetic virtual proteomes contain protein sequence architectures that provide a more biologically grounded framework than random sequence simulations.

While WT proteomes occupied a constrained elemental composition space with low centroid dispersion (median Euclidean distance = 0.37) and limited pairwise variation (median distance = 0.47; range = 0.0006–2.72), pLUCA monsters showed substantially larger deviations (centroid median distance = 1.23; pairwise median distance = 1.30; range = 0.52–5.11), although they still partially overlapped with WT. In contrast, Trifonov monsters deviated even further (centroid median distance = 2.80; pairwise median distance = 2.83; range = 1.53–5.66), corresponding to approximately 3-fold and 8-fold increases relative to WT proteomes, respectively. **(Supplementary information 08, 09)**.

To quantify directional elemental shifts, we calculated per-genome differences relative to WT (delta = Δ) **(Fig. 3A)**. pLUCA monsters consistently showed reduced C and N (median ΔC = −0.63; ΔN = −0.69) with increased O and S (ΔO = 0.57; ΔS = 0.26), whereas Trifonov monsters exhibited larger shifts overall, including stronger reductions in carbon, nitrogen, and sulfur together with marked oxygen enrichment (median ΔC = −0.98; ΔN = −1.20; ΔO = 2.30; ΔS = −0.21) **(Supplementary information 10)**. To test whether these compositional shifts depended on amino acid identity rather than alphabet size alone, we compared amino acid depletion and elemental shifts relative to WT using Spearman correlations. Although expected relationships were identified, including sulfur depletion associated with cysteine (Cys) and methionine (Met) loss, many amino acid-element relationships differed substantially or reversed sign between alphabets, including carbon with isoleucine (Ile) ($\rho = 0.85$ in pLUCA, −0.89 in Trifonov), hydrogen with arginine (Arg) (−0.88, 0.75) and phenylalanine (Phe) (0.79, −0.74), and oxygen with threonine (Thr) (0.62, −0.85) **(Fig. 3 B) (Supplementary information 11)**, indicating that reduced alphabets reorganize elemental composition differently depending on amino acid identity. To assess whether monster proteomes retained detectable homologous signal rather than collapsing into effectively random sequences, we performed BLASTp comparisons against matched WT. Trifonov monsters retained substantially stronger similarity to WT proteins than pLUCA monsters, recovering more WT hits (67.1% vs 34.7%) together with higher alignment coverage, identity, and bitscores **(Fig. 3 D) (Supplementary information 12)**. Despite strongly perturbing elemental composition, both monster proteomes retained substantial detectable primary sequence similarity to modern proteins. Overall, primordial-like alphabets generated elemental compositions divergent from the constrained regime shared by LUCA and extant proteomes, which may suggest that the establishment of the modern proteome elemental profile accompanied earlier and less gradual expansion of the amino acid alphabet than previously proposed.

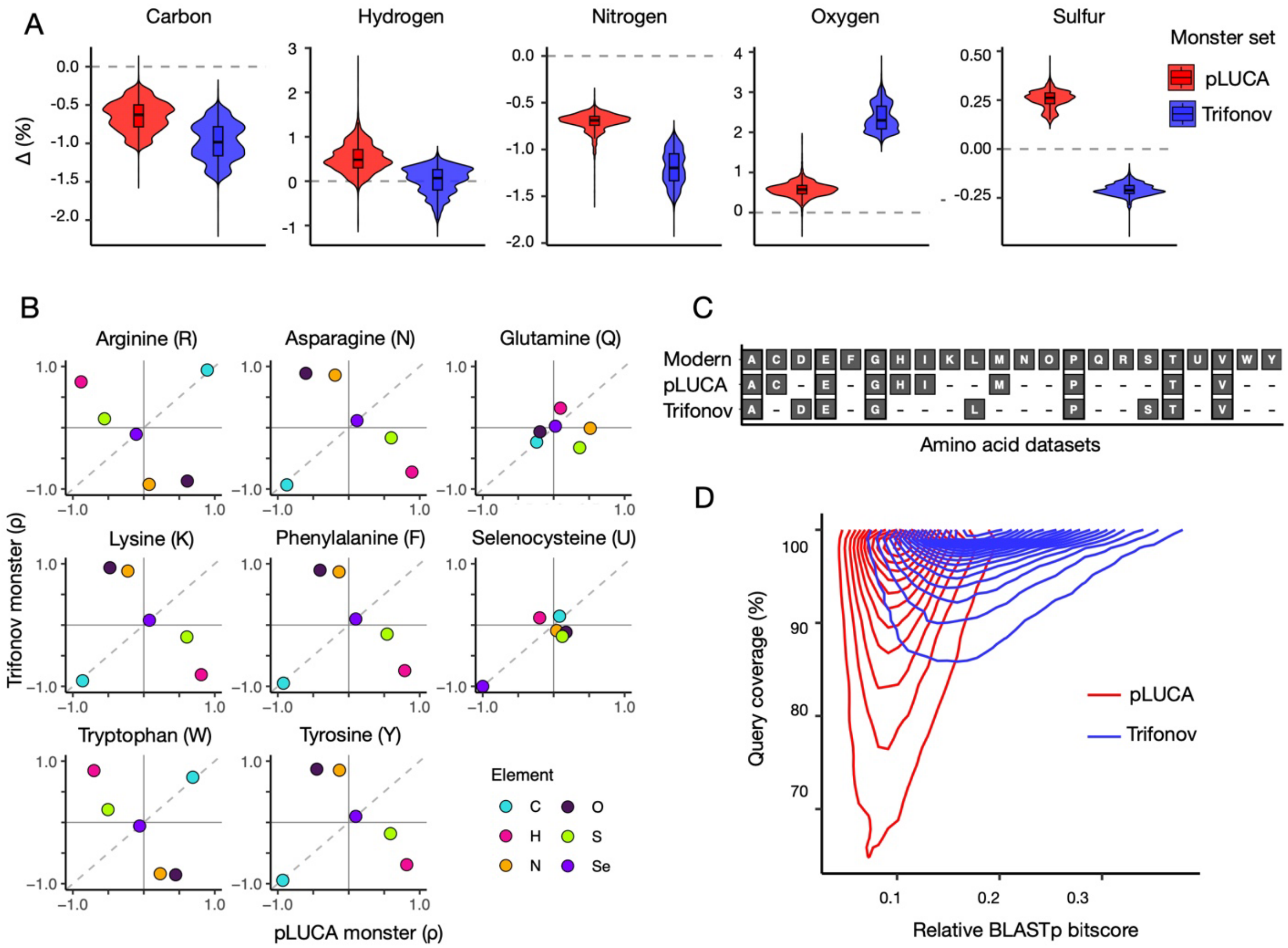

**Figure 3. Reduced primordial-like amino acid alphabets disrupt modern proteome elemental organization.** **(A)** Violin plots showing per-monster proteome elemental composition differences relative to matched wild-type (WT) bacterial proteomes (Δ) for carbon, hydrogen, nitrogen, oxygen, and sulfur in synthetic reduced-alphabet proteomes. Boxplots indicate medians and interquartile ranges. Dashed horizontal lines indicate no difference relative to WT proteomes (Δ = 0). **(B)** Scatterplots comparing Spearman correlations (ρ) between amino acid depletion and elemental shifts relative to WT proteomes for shared amino acids removed in both monster proteomes. Each point corresponds to an individual element. Deviations from the diagonal indicate distinct amino acid–element relationships between pLUCA and Trifonov proteomes, including reversals in correlation direction. **(C)** Amino acid composition of the modern encoded alphabet and reduced hypothetical early alphabets used to generate synthetic pLUCA and Trifonov monster proteomes. Gray boxes indicate retained amino acids, dashes indicate deleted residues, and black square outline is conserved residues. **(D)** Relative BLASTp bitscore versus query coverage for best-hit alignments between reduced-alphabet monster proteomes and matched WT proteins.

## Proteome elemental composition is linked to higher-order structural organization

Finally, because reduced alphabets systematically disrupted the constrained elemental composition characteristic of extant proteomes despite retaining detectable primary sequence conservation, we investigated whether additional constraints emerge from higher-order protein structural organization. Using ESMFold (Lin et al. 2023) predictions from randomly sampled bacterial proteomes (n = 5 per dataset), we quantified α-helices, β-sheets, disorder, and fold confidence relative to elemental composition in WT, pLUCA, and Trifonov monster proteomes.

We found that WT proteomes occupied a compact structure to elemental composition regime characterized by higher α-helix and β-sheet propensities together with lower disorder and consistent elemental abundances. In contrast, both reduced alphabets shifted toward high-disorder regimes associated with reduced C, N and increased O abundances. pLUCA proteomes showed the strongest collapse in α-helix propensity and highest disorder, whereas Trifonov proteomes retained relatively higher ordered structure despite substantial reorganization of element–structure relationships, including complete sulfur depletion **(Fig. 4) (Supplementary information 13)**.

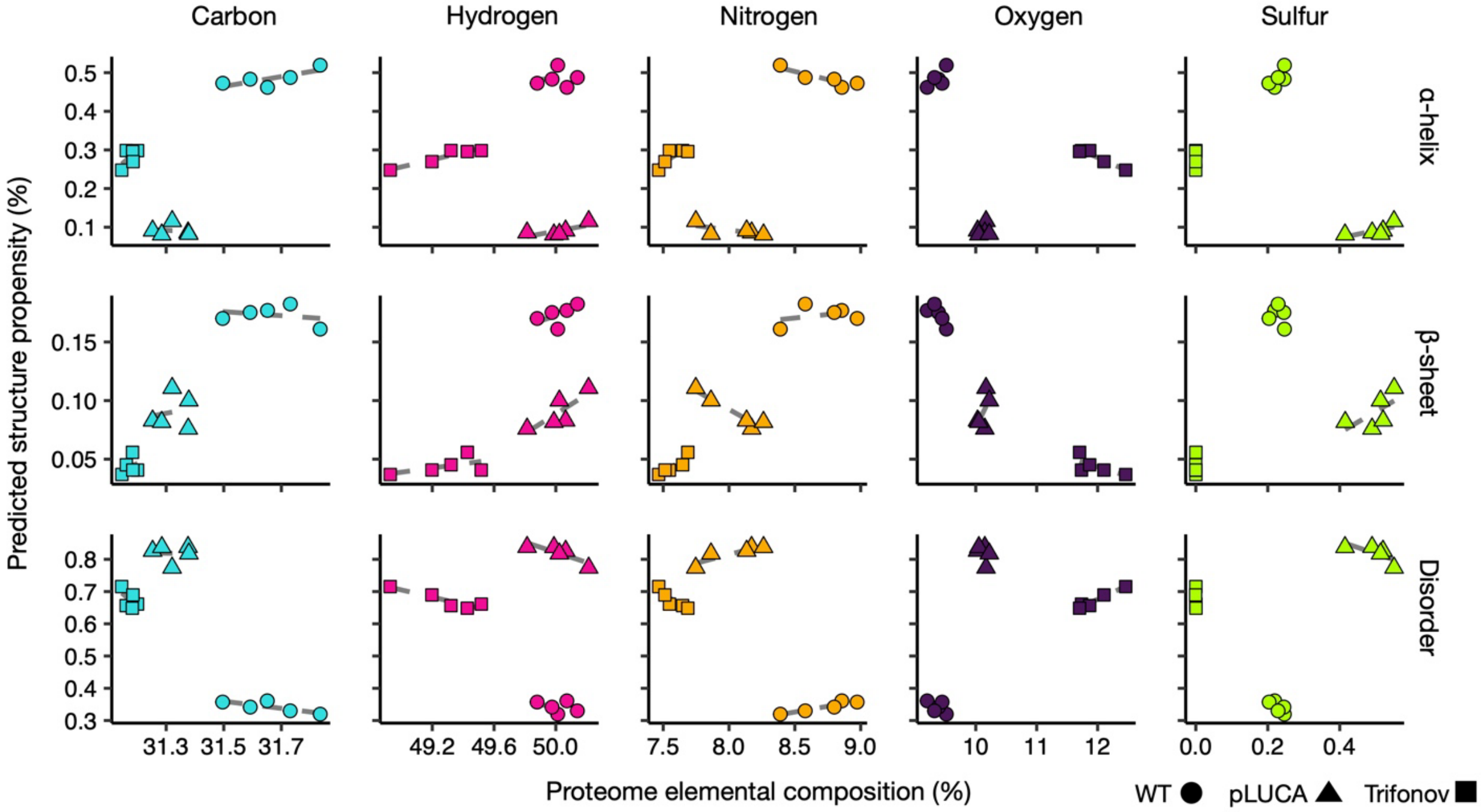

**Figure 4. Reduced primordial-like amino acid alphabets reorganize relationships between proteome elemental composition and predicted protein structural organization.** Scatterplots showing relationships between normalized proteome elemental composition and predicted structural propensities for α-helices, β-sheets, and disorder across WT, pLUCA, and Trifonov proteomes. Each point represents a sampled proteome analyzed using ESMFold predictions. Colors indicate individual elements (carbon, hydrogen, nitrogen, oxygen, and sulfur), whereas point shapes denote proteome datasets (WT, pLUCA, and Trifonov). Dashed gray lines indicate linear regression trends within each dataset and highlight distinct relationships between proteome elemental composition and predicted structural organization.

Spearman correlations showed coordinated element–structure relationships in WT proteomes, including positive associations between nitrogen and disorder ($\rho = 0.9$) together with negative associations between nitrogen and α-helix propensity ($\rho = -0.9$), whereas sulfur showed the opposite relationship with disorder ($\rho = -0.8$). Relative to WT, pLUCA proteomes retained broadly similar correlation structure, whereas Trifonov proteomes exhibited stronger reorganization, including reversal of nitrogen–disorder correlations from strongly positive in WT ($\rho = 0.9$) to maximally negative in Trifonov proteomes ($\rho = -1.0$) **(Supplementary information 14)**. Both monster proteomes showed significantly lower fold confidence than WT proteomes for both pLDDT (Kruskal–Wallis $\chi^2 = 11.18$, $P = 0.0037$) and pTM ($\chi^2 = 12.02$, $P = 0.0025$), with pairwise Wilcoxon tests confirming reduced fold confidence relative to WT (all adjusted $P = 0.0119$) **(Supplementary information 15)**. Together, these results suggest that the constrained elemental composition characteristic of extant proteomes is linked to higher-order protein structural organization.

**Implications for the evolution of the amino acid alphabet and proteome elemental organization**

Here we describe a striking consistency in proteome elemental composition across cellular and viral life. This consistency is not explained by evolutionary relatedness, biological function, or amino acid frequencies and physicochemical properties alone, indicating that proteome elemental organization reflects higher-order biochemical constraints operating across diverse lineages. Similar long-term stability has been proposed for other aspects of cellular chemistry, including microbial stoichiometry (Neveu et al. 2016), inorganic elemental composition in Bacteria and Archaea (Novoselov et al. 2017), and metal usage across billions of years despite major environmental transitions (Schopf et al. 2015). Together, these observations suggest that core intracellular chemistry may be more conserved than external geochemical conditions.

A notable aspect of our results is the elemental convergence between viral and cellular proteomes. Viral proteomes across the major realms occupy largely the same elemental composition space observed in cellular organisms despite the absence of a single viral common ancestor. This con-vergence therefore cannot be explained by shared ancestry, but instead suggests that similar biochemical constraints shape proteome organization whenever proteins evolve from the modern amino acid alphabet.

Several observations further suggest that these constraints originated early in evolution. Despite proposed differences in primordial amino acid usage and environmental chemistry, LUCA-derived proteomes already occupy the constrained elemental composition space observed in modern Bacteria and Archaea. In contrast, reduced primordial-like alphabets systematically generated alternative elemental regimes outside the modern range, even when embedded within realistic modern proteome architectures. Importantly, these synthetic proteomes did not collapse into random sequences but retained substantial detectable homology while exhibiting systematic disruptions in elemental composition, fold confidence, and structure–element relationships. Different reduced alphabets also produced distinct and non-random elemental reorganizations, indicating that proteome elemental organization depends not only on amino acid frequencies, but also on overall alphabet composition and higher-order structural constraints linked to protein foldability and stability (Makarov et al. 2023). In this framework, elemental composition may represent an additional systems-level constraint acting alongside foldability, catalytic function, and chemical availability during the early evolution of the encoded amino acid alphabet. Determining whether the encoded alphabet is exceptional in its ability to stabilize this elemental regime will require future studies using chemically plausible alternative alphabets and evolutionarily realistic models of primordial proteome evolution. Together, our findings raise the possibility that stabilization of proteome-wide elemental organization contributed to the evolutionary selection of the encoded amino acid alphabet. More broadly, constrained proteome elemental composition may represent a fundamental organizational property of biological proteomes that emerged early in evolution and remained consistent across the diversification of cellular and viral life.

## Materials and Methods

### Proteome data sets

Proteomes (full predicted proteins from genomes) were downloaded from two sources: NCBI for cellular proteomes (https://www.ncbi.nlm.nih.gov/datasets/genome) and Joint Genome Institute (JGI) Microbial Genomes/Virus (IMG/VR v4) for viral proteomes (https://genome.jgi.doe.gov/portal/pages/dynamicOrganismDownload.jsf?organism=IMG_VR).

For cellular proteomes the dataset was filtered to include reference genomes with complete assem-bly level and curation from RefSeq, and excluded atypical genomes metagenome assembly genomes (MAGs) and data from large multi-isolate projects. Although this conservative approach reduced taxonomic representation, it ensured high-quality and standardized proteomes for reliable calculation of proteome-wide elemental composition, amino acid frequencies, and protein fold properties. After filtration, and complete taxonomic lineage information mapping from https://www.ncbi.nlm.nih.gov/taxonomy, our final cellular dataset comprises 6,964 proteomes in total, from Archaea (n = 424), Bacteria (n = 6,454), and Eukaryota (n = 86) domains. For proteome metadata and accession identities see **Supplementary table 01**.

For viral proteomes, we included both reference genomes and high-quality viral metagenome-assembled genomes (MAGs) with estimated completeness between 90–100%. In contrast to cellular organisms, much of currently available viral diversity is represented by metagenomically recovered genomes rather than curated isolate references; therefore, restricting analyses to reference genomes alone would substantially reduce representation of known viral sequence diversity. Although this broader inclusion strategy increased dataset size and taxonomic representation compared to cellular proteomes, stringent completeness thresholds were applied to retain high-confidence suitable for proteome-wide compositional analyses. After filtering, the final viral dataset comprised 488,025 proteomes representing the viral realms Adnaviria (n = 37), Duplodnaviria (n = 359,809), Monodnaviria (n = 34,366), Riboviria (n = 90,027), Ribozyviria (n = 16), and Varidnaviria (n = 3,770). Metadata and accession identifiers are provided in **Supplementary table 01**. Taxonomic partitioning at fasta sequence level by viral realm was conducted with SeqKit v2.8.0 (Shen et al. 2016).

**Random proteome datasets**

Random proteomes (n = 1000) were generated with parameters derived from the average number of proteins and the observed minimum and maximum protein lengths across all bacterial proteomes to approximate realistic proteome architecture. Each random proteome consisted in 3,500 random protein sequences ranging from 30 to 4,200 amino acids in length, with randomly sample 22 encoded amino acids using a custom Python script (**generate_random_sequences_commented.py**).

**Calculations of elemental composition, amino acid frequencies and physical chemical properties of cellular and viral proteomes**

Individual protein FASTA sequences from each proteome were analyzed. Amino acids were converted to their corresponding molecular formulas (MFs) retrieved from PubChem (https://pubchem.ncbi.nlm.nih.gov) (Supplementary Table 02). Standard amino acid formulas were adjusted to account for water loss during peptide-bond formation while retaining the constant peptide backbone across residues. Elemental counts were calculated using a custom Python script **(get_atom_count_amino_acids_polymers_v1.py).** For each protein, elemental counts of C, N, H, O, S, and Se were calculated and subsequently summed at the proteome level. Elemental composition was then normalized for each proteome by dividing the total count of each element by the total number of atoms across the proteome **(Combined_protein_aa_atom_count_polymers.tsv Supplementary Table 2).** Amino acid frequencies and physicochemical properties were calculated using custom Python scripts (**script: get_aa_properties_v1.py**). Frequencies were determined by counting the occurrence of each of the 22 amino acids in each protein and normalizing by total protein length. For physicochemical properties, the relative frequencies of amino acids classified as aliphatic, aromatic, non-polar, polar, charged, basic, and acidic were calculated for each protein **(Combined_protein_aa_atom_count_polymers_percentages.tsv Supplementary Table 3).** Through most of the manuscript, amino acids are represented by their 3 letter codes: Ala (Alanine), Arg (Arginine), Asn (Asparagine), Asp (Aspartic acid), Cys (Cysteine), Gln (Glutamine), Glu (Glutamic acid), Gly (Glycine), His (Histidine), Ile (Isoleucine), Leu (Leucine), Lys (Lysine), Met (Methionine), Phe (Phenylalanine), Pro (Proline), Pyl (Pyrrolysine), Sec (Selenocysteine), Ser (Serine), Thr (Threonine), Trp (Tryptophan), Tyr (Tyrosine), and Val (Valine).

**LUCA-reconstructed proteome datasets**

We retrieved 969 ancestrally reconstructed LUCA proteins (ASR) from Wehbi et al. (2024), provided directly by the authors. To capture the full sequence diversity underpinning these reconstructions, we additionally collected 5,711,432 extant sequences from the 969 Pfam protein families used in their ASR pipeline (https://github.com/sawsanwehbi/Pfam-age-classification and

https://figshare.com/articles/dataset/Prokaryotic_Pfams_zip/25599819/1?file=45642522). A second dataset was obtained from Moody et al. (2024). From their Supplementary Data 1, consisting of 399 KEGG KOs (101,484 protein sequences) with a probability of 1 in either the "Probable_and_sampling_threshold_met" or "Possible_and_sampling_threshold_met" columns, from their Figshare repository (https://figshare.com/articles/dataset/The_nature_of_the_Last_Universal_Common_Ancestor_and_its_impact_on_the_early_Earth_system/24428659?file=49101151). A last LUCA dataset incorporated consists of 357 COG families (732,955 sequences) curated by Crapitto et al. (2022) (https://datadryad.org/dataset/doi:10.5061/dryad.5hqbzkh7s). Although LUCA is estimated to have encoded fewer than ~1,000 proteins, all orthologous descendant sequences were retained to capture the distribution of elemental profiles across all putative LUCA-associated protein families.

**Computational modification of bacterial proteomes into reduced-alphabet synthetic "monster" proteomes**

To generate synthetic proteomes using a virtual computational mutagenesis approach we used all bacterial proteome dataset (n = 6,454) (wild type = WT) and deleted the late-entering amino acids from all proteins according to two hypotheses for the evolutionary ordering of the amino acid alphabet. For one modification we classified the amino acids D, F, K, L, N, Q, R, S, W, and Y as pre-LUCA (pLUCA) according to Wehbi et al. (2024). For the Trifonov model, amino acids appearing after the nine earliest Miller-supported residues (G, A, V, D, P, S, E, L, T) were classified as late additions according to a consensus chronology (Trifonov, 2000). Modifications were done with custom Python scripts (**get modified proteins.py**).

**Protein folding and structural analyses of synthetic "monster" proteomes**

We randomly selected five bacterial proteomes (WT), to reduce computational burden, and their matched modified monster proteomes (pLUCA and Trifonov), to generate whole-proteome structural profile using ESMFold implemented through the Transformers Python package with publicly available ESM-2 weights (Li et al. 2023). ESMFold was selected because it does not require multiple-sequence alignments or external databases, making it suitable for synthetic monster proteins with low sequence similarity and expected low template coverage (Manfredi et al. 2025). From structural predictions we calculate the Predicted structures included predicted local distance difference test (pLDDT), predicted template modeling (pTM), and atomic coordinates.

Protein folding calculations were performed on the SOL HPC system (Jennewein et al. 2023) using Gaudi2 accelerators with the optimum-habana framework. Downstream structural analyses were performed using Biotite (Kunzmann et al. 2023) and DSSP v4.6 (Hekkelman et al. 2025). Secondary structures were classified using DSSP annotations and grouped into sheets (B, E), helices (H, G, I, P), and disorder (C, S, T), where H represents α-helices, G $3_{10}$-helices, I π-helices, P κ-helices (poly-proline II helices), B isolated β-bridges, E extended β-strands, T hydrogen-bonded turns, and S bends. Relationships between elemental composition and predicted structural properties were evaluated using Spearman correlations. To assess structural stability, 20 randomly selected proteins were analyzed using all-atom molecular dynamics simulations in OpenMM (Eastman et al. 2023). Code for folding and analyses is available at https://github.com/ELIFE-ASU/monsterproteinstability.

**Statistical tests**

**Variation and distribution of proteome elemental composition**

Proteome size distributions were evaluated using empirical cumulative distribution functions (EC-DFs) for the number of proteins and total atomic content per proteome. For each cellular domain and viral realm, proteome-wide elemental composition (C, H, N, O, S, and Se percentages) was summarized using median, interquartile range (IQR), mean, standard deviation, and minimum–maximum values. Median and IQR were used as robust descriptors of central tendency and dispersion. Pairwise differences in elemental composition between groups were assessed using the two-sample Kolmogorov–Smirnov (KS) statistic (D), which measures the maximum difference between cumulative distributions and ranges from 0 (identical distributions) to 1 (complete separation). Due to large sample sizes, emphasis was placed on KS effect sizes rather than p-values.

**Multivariate analyses of elemental composition**

Multivariate differences in elemental composition among groups were evaluated using permutational multivariate analysis of variance (PERMANOVA) with the adonis2 function from vegan (Oksanen et al. 2026) in R (v4.2.1). Analyses were performed on Euclidean distances calculated from elemental percentages. To reduce computational burden while maintaining balanced representation, datasets were randomly subsampled to up to 500 proteomes per group using a fixed seed for reproducibility. Significance was assessed using 999 permutations, and the proportion of variance explained by group identity ($R^2$) was used as the primary effect-size metric. Principal component analysis (PCA) was also performed to visualize multivariate variation in elemental composition across genomes.

**Amino acid frequency and property analyses**

Relative divergence among proteomes was quantified using an inverted ratio (IR) metric calculated from normalized pairwise differences between proteome feature vectors. IR values were computed independently for elemental composition, amino acid frequencies, and physicochemical properties. Lower IR values indicate greater conservation, whereas higher values indicate greater divergence between proteomes. To evaluate multivariate constraint across proteomes, we quantified dispersion relative to group centroids using Euclidean distances calculated from normalized feature vectors. Dispersion analyses were performed independently for elemental composition, amino acid frequencies, and amino acid physicochemical properties. Lower distances indicate tighter clustering and greater conservation within groups. Proteome amino acid frequencies were compared across groups using pairwise KS statistics. For each pair of groups, distributions of amino acid frequencies were compared independently for all encoded amino acids using proteome-level frequency values, and KS statistics (D) were extracted as measures of distributional difference. Spearman rank correlations were calculated between proteome-level elemental composition (C, H, N, O, S, Se) and amino acid frequencies. Amino acid frequencies were computed per proteome by summing residues across all proteins and normalizing by total amino acid counts. Correlations were calculated across all genomes (pooled) and independently within each group. Correlation strength was evaluated using the absolute value of Spearman's $\rho$. Associations were classified as negligible ($|\rho| < 0.1$), weak (0.1–0.39), moderate (0.4–0.69), strong (0.7–0.89), or very strong ($\geq$0.9). To assess proteome correlation global trends, median and mean absolute correlations were calculated across all elements–amino acid pairs. Recurrent strong associations were identified by counting element–amino acid pairs with $|\rho| \geq 0.7$ across multiple groups.

**Multivariate distance analyses of monster proteomes**

Euclidean distances were calculated from normalized elemental composition vectors (C, H, N, O, S, and Se percentages). A wild-type (WT) bacterial centroid was defined as the mean elemental composition across all WT bacterial proteomes, and Euclidean distances from each WT and monster proteome to this centroid were calculated to quantify multivariate displacement. Pairwise Euclidean distances among WT proteomes and between monster and WT proteomes were also calculated to compare compositional divergence and overlap within elemental composition space.

**Elemental shift and amino acid depletion analyses in monster proteomes**

For each genome, elemental compositional shifts in monster proteomes were calculated relative to matched wild-type (WT) bacterial proteomes as Δ = monster percentage − WT percentage for C, H, N, O, S, and Se. Median Δ values, interquartile ranges, and the percentage of genomes showing increased or decreased elemental composition were summarized for each monster dataset. Distributional differences between WT and monster proteomes were also assessed for each element using two-sample KS statistics. To evaluate whether elemental shifts depended on the identity of depleted amino acids rather than reduced alphabet size alone, amino acid depletion values were calculated as the difference between WT and monster proteome amino acid frequencies. Spearman correlations between amino acid depletion values and elemental Δ values were then calculated independently for pLUCA and Trifonov monster proteomes using matched genomes shared among WT and monster datasets.

**Sequence similarity analyses of monster proteomes**

Protein sequence similarity between proteomes was evaluated using DIAMOND BLASTP (Buchfink et al. 2015). pLUCA and Trifonov monster proteomes were compared against their matched WT bacterial proteomes, and WT proteomes were also compared against themselves as controls. Searches were performed using the parameters --ultra-sensitive, --evalue 1e-5, and --max-target-seqs 5. Best high-scoring segment pairs (HSPs) were retained for downstream analyses. Alignment coverage and bitscore values were calculated for matched proteins. To normalize sequence similarity across proteins of different lengths and compositions, we calculated relative bitscores by dividing the monster-versus-WT bitscore by the corresponding WT self-alignment bitscore for each protein. Lower relative bitscores indicate greater sequence divergence from the WT proteome.

### Protein fold and structural property analyses

Protein fold analyses were performed using proteome-level structural predictions from wild-type (WT), pLUCA, and Trifonov monster proteomes. Predicted secondary structure proportions (alpha helix, beta sheet, and disorder), pLDDT, and pTM values were extracted from protein-level fold prediction datasets and summarized at the proteome level by averaging values across all proteins within each proteome. Relationships between elemental composition and predicted structural properties were evaluated using Spearman correlations independently for WT, pLUCA, and Trifonov datasets. To compare global fold metrics among datasets, proteome-level mean pLDDT and pTM values were analyzed using Kruskal–Wallis tests followed by pairwise Wilcoxon rank-sum tests with Benjamini–Hochberg correction for multiple comparisons.

### Data visualization

Data visualization and statistical plotting to generate figures were performed in R using ggplot2 together with the packages dplyr (Wickham et al. 2026a), tidyr (Wickham et al. 2026b), ggridges (Wilke et al. 2025), vegan (Oksanen et al. 2026), and gridExtra (Auguie and Antonov, 2017).

### Acknowledgements

We thank Sawsan Wehbi for providing LUCA ancestral sequence reconstruction data and guidance in parsing the dataset; Gil Speyer and Arizona State University Research Computing for assistance with computational resources; We thank Hypatia Meraviglia, Veronica Mierzejewski and Pilar Vergeli for discussions during early stages of the project, as well all members of the Emergence Lab. We also thank all authors that generate the primary data (sequences and metadata) used in this work. Code development, manuscript grammar editing, and refinement were assisted using ChatGPT (v4–5.5) and Gemini (v3 Flash). All scientific hypothesis, interpretations, analyses, and conclusions were performed and verified by the authors. This work was supported by NASA Early Career Collaboration Award, and Interdisciplinary Consortia for Astrobiology Research (ICAR) grant 80NSSC21K1402, and the Schmidt Sciences Foundation through a Science Polymath Fellowship awarded to SIW.

### Author contributions

Conceptualization, methodology design and manuscript writing and editing were performed by L.F.B. and S.I.W. Funding acquisition and project management was by S.I.W. Protein folding analyses were performed by L.S. Bioinformatics, statistical analyses, data visualization and scripting were performed by L.F.B. All authors reviewed the final version of this manuscript.

Statistical analysis